\documentclass[prb,superscriptaddress,twocolumn,showpacs]{revtex4}
\usepackage{amsmath}
\usepackage{bm}
\usepackage{amssymb}
\usepackage{graphicx}
\def\bx{{\mbox{\boldmath $x$}}}

\begin{document}

\title{Time-dependent treatment of tunneling and Time's Arrow   problem}

\author{Shmuel Gurvitz}
\email{shmuel.gurvitz@weizmann.ac.il}

\affiliation{Department of Particle Physics and Astrophysics\\  Weizmann Institute of
Science, Rehovot 76100, Israel}
\date{\today}

\pacs{72.10.-d, 72.10.-Bg, 72.20.Dp}

\begin{abstract}
New time-dependent treatment of tunneling from localized state to continuum is proposed. It does not use the Laplace transform (Green's function's method) and can be applied for time-dependent potentials, as well. This approach results in simple expressions describing dynamics of tunneling to Markovian and non-Markovian reservoirs in the time-interval $-\infty<t<\infty$. It can provide a new outlook for tunneling in the negative time region, illuminating the origin of the time's arrow problem in quantum mechanics. We also concentrate on singularity at $t=0$, which affects the  perturbative expansion of the evolution operator. In addition, the decay to continuum in periodically modulated tunneling Hamiltonian is investigated. Using our results, we extend the Tien-Gordon approach for periodically driven transport, to oscillating tunneling barriers.
\end{abstract}
\maketitle

\section{Introduction}

In this paper we perform detailed analysis of tunneling to continuum, by solving time-dependent Schr\"odinger equation in the time-domain. It is based on a new method, proposed for electron transport in mesoscopic systems \cite{gur,GAE}. The method allows us to find simple analytical expressions for tunneling to Markovian and non-Markovian reservoirs, where the Hamiltonian is  time-dependent.

The present paper discusses two subjects. The first one deals with  tunneling dynamics to reservoirs of infinite and finite band-width. A special attention is paid to singularity in the time-evolution at $t=0$ and to the time's arrow problem. The latter is illuminated by our treatment, which does not impose forward propagation of the wave-function in time, referred as ``causality''. (This is different from the standard Green's function technique based on the Laplace transform.)

The causality restriction is not necessary in our approach, since any solution of the first order differential equation (as the time-dependent Schr\"odinger equation) is uniquely defined in the whole interval $-\infty<t<\infty$ by a value of the wave-function at $t=0$. It implies a somehow different outlook on motion at negative time, than usually accepted. For instance, for particle tunneling to continuum from a quantum well, the wave-function at $t<0$ would describe formation of the localized state in the well from the extended states in continuum, (and not a motion in the opposite direction of time).

The second part deals with a periodically modulated time-dependent Hamiltonian. The analytical expressions for the case of oscillating energy level and the tunneling barrier, obtained in the present paper would allow us to find a not-trivial extension of the Tien-Gordon treatment to electron transport through periodically oscillating energy levels \cite{tien}, to periodically oscillating barriers.

The plan of this papers is as follows: Sec.~\ref{sec2} starts with a general description of the system. Sec.~\ref{sec2a} introduces standard treatment of tunneling for the time-independent Hamiltonian. Sec.~\ref{sec2b} presents our method for the time-dependent Hamiltonian, where general expressions for occupation of the well (survival probability) as a function of time and energy spectrum of the tunneling particle in the reservoir are obtained. Sec.~\ref{sec2c} considers wide-band limit (Markovian reservoir). Sec.~\ref{sec3} analyzes singularity in the survival probability for Markovian and non-Markovian reservoirs in relation with the Schre\"odinger evolution. Sec.~\ref{sec4} discusses the time-arrow problem in quantum mechanics, in connection with our approach. Secs.~\ref{sec5} and \ref{sec6} consider tunneling from the oscillating energy level and through the oscillating barrier. An extension of the Tien-Gordon treatment to oscillating barriers is proposed. Sec.~\ref{sec7} is devoted to discussion and summary.

\section{Quantum tunneling\label{sec2}}

\begin{figure}[h]
\includegraphics[width=6cm]{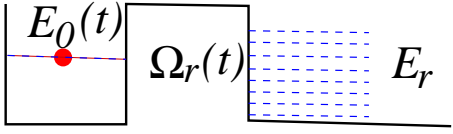}
\caption{(Color online) Tunneling of a particle to continuum, where the energy level $E_0$ and the coupling energy are time-dependent.}
\label{fig1}
\end{figure}

Consider time-evolution of a particle, localized at $t=0$ inside a quantum well at the state  $|0\rangle$. The latter  coupled to extended states $|r\rangle$ of a reservoir, Fig.~\ref{fig1}. The system is described by the following time-dependent Hamiltonian $H(t)=H_0(t)+H_T(t)$, where
\begin{subequations}
\label{a1}
\begin{align}
H_0(t)&=E_0(t)|0\rangle\langle 0|+\sum_rE_r|r\rangle\langle r|\label{a1a}\\
H_T(t)&=\sum_r\Omega_r(t)\big(|r\rangle\langle 0|+|0\rangle\langle r|\big)
\label{a1b}
\end{align}
\end{subequations}
The reservoir levels are time-independent.

Usually, reservoir with a finite band-width $W$ is represented by a semi-infinite lead, consists of periodic one-dimensional chain of $N$-coupled quantum wells with the nearest-neighbor coupling $W/2$, corresponding to
\begin{align}
E_r^{}=W\cos \left({r\pi\over N+1}\right),~~{\rm for}~~r=1,\ldots ,N\, .
\label{tbham1}
\end{align}
For the time-independent Hamiltonian, $\Omega_r(t)=\Omega (E_r)$, the tunneling coupling is given by \cite{gur1}
\begin{align}
\Omega(E_r)=\sqrt{{\Gamma W \over 2\pi(N+1)}}\sqrt{1-{E_r^2\over W^2}}
\label{tbham2}
\end{align}
where $\Gamma$ is a width of the level $E_0$ in the wide-band (Markovian) limit, $W\to\infty$.

When we consider the time-dependence of tunneling couplings $\Omega_r(t)$, we assume that it is generated by the  time-dependent barrier. In this case we use
\begin{align}
\Omega_r(t)\equiv\Omega(E_r)\, w(t)\, ,
\label{omt}
\end{align}
so that $w(t)$ accounts variation of the barrier penetration with time.

The particle's wave function can be written as
\begin{align}
|\Psi (t)\rangle=b_0(t)|0\rangle +\sum_rb_r(t)|r\rangle \, ,
\label{a3}
\end{align}
where he probability amplitudes $b_0(t),\, b_r(t)$ are obtained from the Schr\"odiger equation,
\begin{align}
i\partial_t |\Psi (t)\rangle =H(t)|\Psi (t)\rangle
\label{sch}
\end{align}
subjected to condition $b_0(0)=1$ and $b_r(0)=0$. Substituting Eq.~(\ref{a3}) into Eq.~(\ref{sch}), we find
\begin{subequations}
\label{a2}
\begin{align}
&i\dot b_0(t)=E_0(t)b_0(t)+\sum_r\Omega(E_r)w(t)b_r(t)
\label{a2a}\\
&i\dot b_r(t)=E_r b_r(t)+\Omega(E_r)w(t)b_0(t)
\label{a2b}
\end{align}
\end{subequations}
Solving Eqs.~(\ref{a2}), we obtain the amplitudes $b_{0,r}(t)$, and therefore all information about time-evolution of the system.

\subsection{Laplace transform (Green's function method)\label{sec2a}}

Consider first the time-independent Hamiltonian, $H_0(t)=H_0$ and $H_T(t)=H_T$, corresponding to $E_0(t)=E_0$ and $w(t)=1$, Eq.~(\ref{omt}). In this case the time-dependent Schr\"odinger equation is usually solved by applying the Laplace transform,
\begin{align}
|\tilde\Psi(E)\rangle=\int_0^\infty |\Psi(t)\rangle e^{iEt}dt\equiv \tilde b_0(E)|0\rangle +\sum_r\tilde b_r(E)|r\rangle
\label{laplace}
\end{align}
where the Laplace variable $E$ (``energy'') must be taken above the real axis, $E\to E+i0$ for providing convergence of integral for $t>0$ (causality).

Applying the Laplace transform to the Schr\"odinger equation (\ref{sch}), we can rewrite it as
\begin{align}
|\tilde\Psi (E)\rangle =iG_0(E)|\Psi (0)\rangle +G_0(E)H_T|\tilde\Psi (E)\rangle
\label{lwf}
\end{align}
where
\begin{align}
G_0(E)={1\over E-H_0}\equiv {|0\rangle\langle 0|\over E-E_0}+\sum_r{|r\rangle\langle r|\over E-E_r}
\label{grf}
\end{align}
is the Green's function. Solving Eq.~(\ref{lwf}) we find
\begin{subequations}
\label{twf}
\begin{align}
&\tilde b_0(E)={i\over E-E_0-\sum_r{\Omega^2(E_r)\over E-E_r}}\label{twfa}\\[5pt]
&\tilde b_r(E) ={\Omega(E_r)\over E-E_r}b_0(E)
\label{twfb}
\end{align}
\end{subequations}

In the continuous limit ($N\to\infty$) one can replace $\sum_r\to\int\rho(E_r)dE_r$, where $\rho (E_r)$ is the density of states, $\rho^{-1} (E_r)=dE_r/dr$. Note, that $N$ is cancel out in the reservoir spectral-density  function $S(E_r)=\Omega^2(E_r)\rho(E_r)$, that reads
\begin{align}
S(E_r)={\Gamma\over 2\pi}\sqrt{1-{E_r^2\over W^2}}\, .
\label{fb}
\end{align}
Despite its simple form, the spectral-density function~(\ref{fb}) does not allow us to obtain simple analytical expressions for the amplitudes $\tilde b(E)$ Eqs.~(\ref{twf}) in the continuous limit. However, instead of Eq.~(\ref{fb}) we can use the Lorentzian form for the finite band-width spectral function,
\begin{align}
S(E_r)
={\Gamma\over2\pi}{\Lambda^2\over E_r^2+\Lambda^2}
\label{lor}
\end{align}
where $\Lambda=\sqrt{2}\,W$ provides the same curvature at the band center as  Eq.~(\ref{fb}). This is enough to make the Lorentzian~(\ref{lor}) a very good approximation for a finite range spectral function~(\ref{fb}), as demonstrated in Ref.~[\onlinecite{gur1}].

Consider first the Markovian (wide-band) limit, $\Lambda\to\infty$. Then the tunneling coupling and density of states become energy independent,  $\Omega(E_r)\equiv\Omega$, $\rho(E_r)\equiv\rho$, so that $S(E_r)\to\Gamma/(2\pi)$ and therefore $\int_{-\infty}^\infty S(E_r)/(E-E_r)dE_r\to-i\Gamma/2$. As a result
\begin{subequations}
\label{lapamp}
\begin{align}
\tilde b_0(E)&={i\over E-E_0+i{\Gamma\over2}}\label{lapampa}\\
\tilde b_r(E)&={i\,\Omega \over (E-E_r)\big(E-E_0+i{\Gamma\over2}\big)}
\label{lapampb}
\end{align}
\end{subequations}

The time-dependent amplitudes $b(t)$ are obtained from $\tilde b(E)$ via the inverse Laplace transform,
\begin{subequations}
\label{invlap}
\begin{align}
b_0(t)&=\int\limits_{-\infty}^\infty \tilde b_0(E)e^{-iEt}{dE\over 2\pi}=e^{-iE_0t-{\Gamma\over2}t}\label{invlapa}\\
b_r(t)&={\Omega\,e^{-iE_rt}\over E_r-E_0+i{\Gamma\over2}}\left[1-e^{i\big(E_r-E_0
+i{\Gamma\over2}\big)t}
\right]
\label{invlapb}
\end{align}
\end{subequations}
Then the probabilities of finding the particle inside the well (survival probability), $P_0(t)=|b_0^{}(t)|^2$, or inside the reservoir at the level $E_r$ (line shape), $P_r(t)=|b_r(t)|^2\rho$ are
\begin{subequations}
\label{exx}
\begin{align}
P_0(t)&=e^{-\Gamma t}\label{ex0}\\
P_r(t)&={\Gamma\over 2\pi}\,{1-2\cos [(E_0-E_r)t]\,e^{-{\Gamma\over2}t}+e^{-\Gamma t}\over (E_r-E_0)^2+{\Gamma^2\over4}}
\label{exr}
\end{align}
\end{subequations}

Equation~(\ref{ex0}) displays the exponential decay of $P_0(t)$. As a result, the tunneling particle never returns to its initial state. At first sight, such a behavior is at odds with the Poincare recurrence theorem. In fact, there is no contradiction, since Eqs.~(\ref{exx}) correspond to the continuous limit, $N\to\infty$, where the level spacing of the reservoir's states, $E_{r+1}-E_r\to 0$,  (see Eq.~(\ref{tbham1})), corresponding to reservoir of infinite size. However, for large but finite $N$, the particle returns to its initial state.

\begin{figure}[h]
\includegraphics[width=9.5cm]{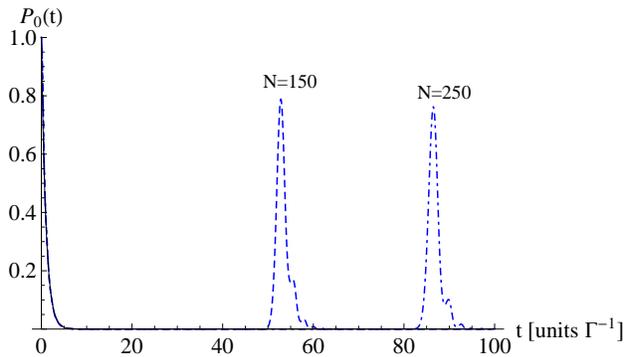}
\caption{(Color online) Survival probability for tunneling to a finite periodic array of $N$ quantum wells for $N=150$ (dashed line) and $N=250$ (dot-dashed line). The exponential decay is shown by solid line.}
\label{figap2}
\end{figure}

This is illustrated in Fig.~\ref{figap2}, which shows the survival probability, $P_0(t)$, obtained by the inverse Laplace transform, Eq.~(\ref{invlapa}) of Eq.~(\ref{twfa}) with $\Omega(E_r)$ given by Eq.~(\ref{tbham2}).
for $W =6\Gamma$, $E_0=\Gamma$ and $N=150$ (dashed line), $N=250$ (dot-dashed line), in comparison with the exponential decay, $P_0(t)=\exp (-\Gamma t)$ (solid line).

It follows from Fig.~\ref{figap2} that for finite-size (finite $N$) reservoir, the survival probability $P_0(t)$ for small times is very close to the exponential decay, but it displays return to the initial state at later times. However, the corresponding return (revival) time increases with $N$, Fig.~\ref{figap2}. Then in the limit, where the reservoir's size increases to infinity the revival time becomes infinitely long. Note that this revival-time phenomenon has has no relation to the time's arrow problem, discussed in  Sec.~\ref{sec4}.

\subsection{Time-dependent Hamiltonian\label{sec2b}}

In the case of time-dependent Hamiltonians, the Laplace-transform method becomes less useful, since it involves convolution integrals.
In addition, it separates the time-domain in two regions: $t\ge 0$ and $t\le 0$, with respect to the initial condition ($t=0$), where the forward propagation in time (causality) is usually considered. In fact, such a separation looks superfluous, since the time-dependent Schr\"odinger equation is the first order differential equation. Therefore the wave-function is uniquely defined in a whole time-interval $-\infty<t<\infty$, by its value at any time ($t_0$) inside the interval, named as ``initial condition''.

In order to solve Eqs.~(\ref{a2}) directly in the time-domain, we  apply a method, derived in Refs.~[\onlinecite{gur,GAE}]. First we resolve Eq.~(\ref{a2b}), thus obtaining
\begin{align}
b_{r}^{}(t)=-i\,e^{-iE_{r}t}\int\limits_{0}^t
\Omega(E_r)w(t') b_{0}^{}(t')e^{iE_{r}t'}dt'
\label{b1}
\end{align}
Substituting Eq.~(\ref{b1}) into Eq.~(\ref{a2a}), we rewrite the latter in the continuous limit as
\begin{align}
i\dot {b}_{0}^{}(t)=E_0(t)\,b_{0}^{}(t)
-i\, w(t)\int\limits_0^t \tilde S(t-t')w(t')b_0^{}(t')dt'
\label{b2}
\end{align}
where
\begin{align}
\tilde S(t-t')=\int\limits_{-\infty}^\infty S(E_r)e^{iE_r(t'-t)}dE_r
\label{g}
\end{align}
Solving the integro-differential equation (\ref{b2}) with respect to
$b_0(t)$ and then substituting it into Eq.~(\ref{b1}), we obtain the survival probability
\begin{align}
P_0(t)=|b_0(t)|^2
\label{endis0}
\end{align}
and the energy distribution of tunneling particle
\begin{align}
P_r(t)=S(E_r)\left|\int\limits_0^t w(t')b_0(t')e^{iE_rt'}dt'\right|^2\, .
\label{endis}
\end{align}

\subsection{Wide-band limit.\label{sec2c}}

Consider tunneling to aMarkovian reservoir by taking the limit $W\to\infty$ in Eq.~(\ref{fb}) (or $\Lambda\to\infty$ in Eq.~(\ref{lor})). Then $\tilde S(E_r)=\Gamma\delta(t-t')$. Substituting it in Eq.~(\ref{b2}), we note that the $\delta$-function contributes at the upper limit of the integral. This requires us to carry out the integration more carefully, by using an explicit representation of the $\delta$-function. For instance
\begin{align}
&\int\limits_0^t\delta (t'-t)dt'=\lim_{\eta\,\to 0}\int\limits_0^t
{1\over\pi}{\eta dt'\over\eta^2+(t'-t)^2}={1\over2}{\rm sgn}(t)
\label{intd}
\end{align}
where sgn$(t)=t/|t|$. Using this result, Eq.~(\ref{b2}) can be written as
\begin{align}
\dot {b}_{0}^{}(t)=
\left[-iE_0(t)-{\Gamma\, w^2(t)\over2}\,{\rm sgn}(t)\right]
\,b_{0}^{}(t)
\label{b4}
\end{align}
Solving this equation we finally obtain
\begin{align}
b_0^{}(t)=e^{-i{\cal E}_0(t)\,t}
\label{b5}
\end{align}
where
\begin{align}
{\cal E}_0(t)\, t=\int\limits_0^t\Big[E_0(t')-i{\Gamma\, w^2(t') \over2}\,{\rm sgn}(t)\Big]dt'
\label{b6}
\end{align}

Respectively, the probabilities of finding the particle inside the well Eq.~(\ref{endis0}) and the energy-distribution inside the reservoir, Eq.~(\ref{endis}), are
\begin{subequations}
\label{bb8}
\begin{align}
P_0(t)&=\exp\Big(-{\rm sgn}(t)\int\limits_0^t\Gamma \, w^2(t') dt'\Big)\label{b8}\\
P_r(t)&={\Gamma\over 2\pi}\left|\int\limits_0^t w(t')
e^{i\big[E_r-{\cal E}_0(t')\big]t'}dt'\right|^2
\label{b9}
\end{align}
\end{subequations}
Note, that Eqs.~(\ref{b5}), (\ref{b6}) represent solution of the time-dependent Schr\"odinger equation  Eq.(\ref{sch}) in the time-interval $-\infty<t<\infty$.

\section{Finite spectral width and singularity at $t=0$\label{sec3}}

Consider Eq.~({\ref{b8}) in the case of time-independent Hamiltonian, $E_0(t)=E_0$ and $\Omega (t)=\Omega$. One finds
\begin{align}
P_0(t)=e^{-\Gamma |t|}
\label{bex0}
\end{align}
The same time-dependence of the survival probability, as corresponding to resonance and anti-resonance pole contribution, has been discussed in Ref.~[\onlinecite{hatano}].

Thus, Eq.~(\ref{bex0}) displays a singularity (cusp) at $t=0$. In addition,  the probability of decay (for $t>0$) from the well to the reservoir, $1-P_0(t)$, is proportional to $t$ at small times. On first sight it seems to be in odds with the Schr\"odinger evolution that predicts the probability of quantum transitions for short-time is  proportional to $t^2$. Indeed,
\begin{align}
|\Psi (t)\rangle =e^{-iHt}|\Psi (0)\rangle =\Big(1-iHt-{1\over2}H^2t^2+\cdots\Big)|\psi (0)\rangle
\label{expwf}
\end{align}
As a result,
\begin{align}
P_0(t)=|\langle\Psi (t)|\Psi (0)\rangle|^2=1-C\, t^2+\cdots
\label{c0}
\end{align}
where
\begin{align}
C=\langle\Psi (0)|H^2|\Psi (0)\rangle -\langle\Psi (0)|H|\Psi (0)\rangle^2
\label{c1}
\end{align}
Note that the quantum Zeno effect \cite{zeno} is based on Eq.~(\ref{c0}).

The question is how Eq.~(\ref{bex0}) can be reconciled with Eq.~(\ref{c0}) at small $t$. In fact, it is impossible, since expansion (\ref{expwf}) of the evolution operator cannot be used in the case of Markovian reservoirs (wide-band limit) \cite{zeno1}. Indeed, one can easily check that $C\to\infty$ in Eq.~(\ref{c0}). This results in singularity of $P_0(t)$ at $t\to\infty$, which explicitly displayed by a cusp, in Eq.~(\ref{bex0}). However, the situation is different for non-Markovian reservoirs with a finite band-width ($\Lambda$).

Indeed, let us consider a Lorentzian spectral density, $S(E_r)$. Eq.~(\ref{lor}). Substituting it into Eqs.~(\ref{b2}), (\ref{g}), we find
\begin{align}
i\dot {b}_{0}^{}(t)=E_0(t)\,b_{0}^{}(t)-i{\Gamma\Lambda\over2}\,
w(t)\int\limits_0^tw(t')\, e^{-\Lambda |t'-t|}b_0^{}(t')dt'
\label{b66}
\end{align}
This is an integro-differential equation. However, it can be transformed  into the second order differential equation by differentiating both sides of this equation. After some algebra one finds
\begin{align}
&i\ddot b_{0}^{}(t)=\Big[E_0(t)-i\,{\rm sgn}(t)\Lambda +i{\dot w(t) \over w(t)}\Big] \,\dot b_{0}^{}(t)\nonumber\\
&+\Big[\dot E_0(t)
+\Big({\rm sgn}(t)\Lambda -{\dot w(t)\over w(t)}\Big)E_0(t)-i\Lambda {\Gamma\, w^2(t)\over2}\Big] b_0(t)
\label{b77}
\end{align}

For the time-independent Hamiltonian, corresponding to $w(t)=1$ and $\dot E_0(t)=0$, Eq.~(\ref{b77}) can be solved analytically,  thus obtaining
\begin{align}
&b_0^{}(t)=\, e^{-i{E_0\over2}t-i{\Lambda\over2}|t|}
\Big[\cosh\Big({Q|t|\over2}\Big)\nonumber\\[5pt]
&~~~~~~~~~~~~~~~~~+{\Lambda -i\,{\rm sgn}(t)E_0\over Q}\sinh\Big({Q|t|\over2}\Big)\Big]
\label{b79}
\end{align}
where
\begin{align}
Q=\sqrt{\Lambda^2-2\Gamma\Lambda -E_0^2-2i\, {\rm sgn}(t)E_0\Lambda}
\label{b80}
\end{align}

Let us evaluate the survival probability $P_0(t)=|b_0(t)|^2$ for small $t$. Expanding  Eq.~(\ref{b79}) in powers of $t$ we find
\begin{align}
P_0(t)=1-{\Gamma \Lambda\over2}t^2+{\Gamma\Lambda^2\over6}|t|^3
+{\cal O}[t^4]
\label{b81}
\end{align}
Thus for a finite bandwidth, $P_0(t)$ contains no linear in $t$ term in agreement with Eq.~(\ref{c0}). Nevertheless the non-analyticity of $P_0(t)$ in $t$ is still persisting. It appears as a cusp in the cubic term of the expansion (\ref{b81}). This non-analyticity can be also  revealed by direct evaluation of the higher order derivatives of $P_0(t)$, Eq.~(\ref{c0}), diverging at $t=0$.  This calls for precaution in using the short-time expansion of the evolution operator, Eq.~(\ref{expwf}).

Note that for a large band-width ($\Lambda\gg E_0,\Gamma^{-1}$), the quadratic time-dependence of the tunneling probability takes place at very short time-intervals ($t<1/\Lambda$), as follows from Eq.~(\ref{b81}). Then the difference between the Markovian and non-Markovian reservoirs would not be essential, so that the decay becomes almost pure exponential when $t$ is large enough.

Occupation of the well as a function of time is displayed in Fig.~\ref{fig11} for $\Lambda = 4\Gamma$ and  $E_0=\Gamma$ by solid line. A pure exponential decay, Eq.~(\ref{bex0}), for $\Lambda\to\infty$ is shown by dashed line. The latter clearly displays the cusp at $t=0$, in contrast with with a smooth behavior for finite $\Lambda$, corresponding to a non-Markovian reservoir.
\begin{figure}[h]
\includegraphics[width=8cm]{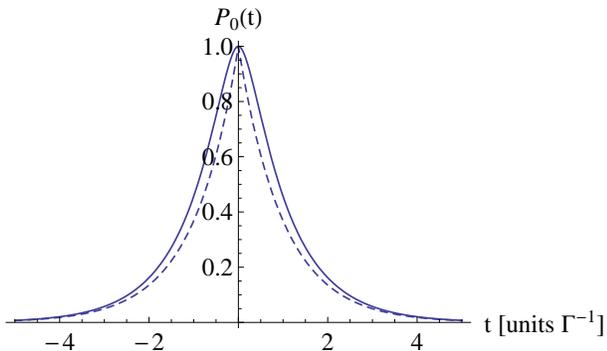}
\caption{(Color online) Probability of finding the particle inside the quantum well as a function of time. The dashed and solid lines corresponds to the Markovian ($\Lambda\to\infty$) and non-Markovian creservoir, Eq.~(\ref{lor}), with $\Lambda = 4\Gamma$, $E_0=\Gamma$.
\label{fig11}}
\end{figure}

\section{Time-arrow problem\label{sec4}}

Consider time-dependent Schr\"odinger equation~(\ref{sch}). The wave-function $|\Psi (t)\rangle$, obtained from this equation is uniquely determined in the interval $-\infty <t< \infty$ by the (initial) condition, $|\Psi (0)\rangle =|0\rangle$, corresponding to a localized state inside the well. For $t>0$ the wave-function describes tunneling from the localized state to extended states of the reservoir.

The question is what process is described by the wave-function $|\Psi (t)\rangle$ for $t<0$. It is quite obvious that it describes formation of the localized state $|0\rangle$ from extended states of the reservoir. Let us elaborate this point on an example of the time-independent Hamiltonian and the Markovian reservoir, $E_0(t)=E_0$, $\Omega_r(t)=\Omega$ in Eqs.~(\ref{a1}), and $\Lambda\to\infty$.

Solving the Schr\"odinger equation (\ref{sch}), (\ref{a2}), we obtain the wave function, $|\Psi (t)\rangle$, given by  Eqs.~(\ref{a3}), with
\begin{subequations}
\label{exx1}
\begin{align}
b_0^{}(t)&=e^{-iE_0t-{\Gamma\over2}|t|}
\label{ex01}\\
b_r^{}(t)&={\Omega\, e^{-iE_rt}\over E_r-E_0+i{\Gamma\over2}}
\left[1-e^{i(E_r-E_0)t-{\Gamma\over2}|t|}\right]\, .
\label{exr1}
\end{align}
\end{subequations}
for $-\infty <t<\infty$. This wave-function displays an appearance of the localized state $|0\rangle$ at $t=0$ as a result of Schr\"odinger evolution from the extended state
\begin{align}
|\varphi_-\rangle =\sum_r{\Omega e^{-i E_rt}\over E_r-E_0+i{\Gamma\over2}}|r\rangle
\label{ininf}
\end{align}
at $t\to -\infty$, followed by decay of the state $|0\rangle$ to the reservoir states at $t\to\infty$.

Actually, the time-evolution of the state $|\varphi_-\rangle$ in the interval $-\infty<t<0$ can be determined by using the time-reversal arguments only. Indeed upon the time-inversion, the Schr\"odinger equation for the wave-function (in configuration space) reads  $i\partial_t \Psi^*(x,-t) =H\Psi^*(x,-t)$ with the initial condition, $\Psi^*(x,0)=\Phi_0(x)$, where $\Phi_0(x)=\langle x|0\rangle$. The latter is an eigen-state of the level $E_0$ inside the well and therefore can be made real by an appropriate gauge. The same (initial) condition defines $\Psi (x,t)$. As a result, $\Psi(x,-t)=\Psi^*(x,t)$ (see  Eqs.~(\ref{b1}), (\ref{b5})).

It follows from Eq.~(\ref{ininf}) that formation of the localized state $|0\rangle$ inside the well takes place from a particular linear superposition of the extended states $|r\rangle$, at $t\to\-\infty$. On first sight, the state $|\varphi_-\rangle$  has a small probability to be recovered. However, it is not the case, since transition from extended to localized state has the same probability as decay from localized to extended state.

In fact, the entire process can be considered as a scattering by quantum well, or as a two-particle scattering, accompanied by a long-lived localized (resonance) state. This takes place from the initial extended state in the reservoir at $t\to -\infty$, followed by a formation of the resonance that subsequently decays back to the reservoir. Such a processes is well observed with no indication on  existence of the time's arrow. (Note that entropy there does not increase and is actually zero, since Eqs.~(\ref{exx1}) describe a pure state \cite{fn1}).

The problem however, arises in many-body systems, where the time-arrow does exist. It is attributed to increase of entropy, so that the less disordered state will be more disordered in future, but not vice versa. However, from time-symmetry arguments one concludes that it was also more disordered in past \cite{legg}. Indeed, let us introduce the wave function $\Psi (\bx,t)$ for an entire system in the time-interval $-\infty<t<\infty$, where $\bx$ denotes the system variables. The total wave-function is uniquely determined by the initial condition,  $\Psi(\bx,0)=\Phi_0(\bx)$. In general, $\Psi^*(\bx,0)=\Phi_0^*(\bx)\not =\Phi_0(\bx)$ (the complex conjugation inverts the sign of momenta).
Applying the time-symmetry one obtains, $\Psi^*(\bx,t)=\Psi(\bx,-t)$. Then from the entropy arguments $\Psi^*(\bx,t\to\infty)$ described the mostly disordered state. Respectively, $\Psi(\bx,t\to -\infty)$ must also be mostly disordered. The same argument can be applied, when the system is in the mixture of pure states, described by the density-matrix.

Of course, the above arguments cannot be applied, if the time cannot be extended to $-\infty$, but is restricted by some value, or if the total system is not isolated. However, one can always expand the system in such a way that it can be considered as well isolated during very long period of time. Then the contradiction between time-reversal symmetry and entropy increase cannot be refuted.

A possible solution of this paradox might be found in quantum measurement theory \cite{ahar,kor}. This implies that the measurement projects the total wave-function of entire system on  eigen-states of a dynamical variable, which can be directly accessible to observer. Such projections in the case of consequitive measurements, break time-reversal symmetry of the Schr\"odinger equation and could lead to time-arrow \cite{kor}. This  happens even in non-selective measurements, when the result of measurement is discarded. The problem is whether the measurement take place when  nobody observes it \cite{gur1}? If yes, the quantum measurement can indeed be relevant for the time-arrow problem, although many complicated questions are needed to be resolved.

\section{Oscillating energy level\label{sec5}}

Equations (\ref{b5})-(\ref{bb8}) and (\ref{b77}) are valid for any time-dependence of $E_0(t)$ and $\Omega_r(t)$ in the Hamiltonian (\ref{a1}). Consider oscillating energy level of the dot
\begin{align}
E_0(t)=E_0-u\sin (\omega t)\, ,
\label{e0t}
\end{align}
while all other parameters in the Hamiltonian are time-independent. In the case of Markovian reservoir ($\Lambda \to\infty$), the amplitude $b_0^{}(t)$ is given by Eq.~(\ref{b5}). One finds $P_0(t)=\exp (-\Gamma |t|)$, Eq.~(\ref{endis0}), so that it is not affected by oscillations of the energy level. This is not surprising, since the total tunneling rate is independent of $E_0$ in the wide-band limit.

However, is not the case for a reservoir with a finite band-width $\Lambda$, Eq.~(\ref{lor}). Indeed, consider Eq.~(\ref{b77}) for $w(t)=1$. Solving it, we find the occupation of quantum well as a function of time. The results are shown in Fig.~\ref{fig3} for $\Lambda =4\Gamma$, $u=3\Gamma$ and two values of the frequency, $\omega =0$ (solid line), $\omega =2\Gamma$ (dashed line, red). The panel (a) corresponds to $E_0=3\Gamma$ and the panel (b) corresponds to $E_0=0$ (aligned with the band-center).
\begin{figure}[h]
\includegraphics[width=8cm]{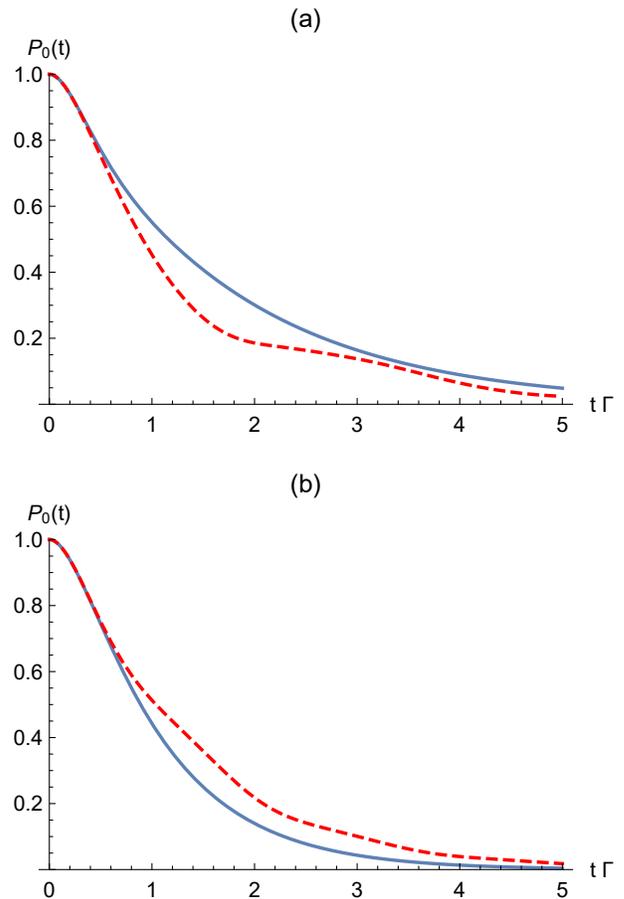}
\caption{(Color online) Probability of survival for oscillating energy level for $\Lambda =4\Gamma$, $u=3\Gamma$, and $\omega =0$ (solid) or $\omega =2\Gamma$ (dashed, red), where (a) $E_0=3\Gamma$, and (b) $E_0=0$.
\label{fig3}}
\end{figure}

One finds from this figure that the oscillations speed up the tunneling if $E_0$ is misaligned with the band-center (a), and slow it down when they are aligned (b). This phenomenon resembles the anti-Zeno and Zeno effects \cite{zeno,azeno} and  can be interpreted in a similar way \cite{zeno2}. Indeed, if the level $E_0$ is not aligned with the reservoir's band-center, Fig.~\ref{fig3}a, the oscillations drive $E_0$ more close to it. As a result, the decay speeds up. However, if $E_0=0$, as in Fig.~\ref{fig3}b, the oscillations drive $E_0$ away from zero. In this case the decay slows down.

A more pronounced effect of the oscillations appears in the energy  spectrum of the tunneling particle and it also takes place for $\Lambda\to\infty$.  Indeed, by using
\begin{align}
\exp \Big[-i\,{u\over\omega}\cos(\omega t)\Big]=\sum_{n=-\infty}^{\infty}
(-i)^nJ_n\Big({u\over\omega}\Big)e^{i\, n\omega t}
\end{align}
where $J_n(x)$ are the Bessel functions of the first kind and substituting this expression into  Eq.~(\ref{b9}) we obtain for the energy spectrum of tunneling particle, $\bar P(E_r)=P_r(t\to\infty )$
\begin{align}
\bar P(E_r)={\Gamma\over 2\pi}\left|\sum_{n=-\infty}^\infty{(-i)^nJ_n\big({u\over \omega}\big)\over E_r-E_0-n\omega +i{\Gamma\over2}}\right|^2
\label{c1}
\end{align}

This result has a simple physical interpretation in terms of the Floquet states and the Tien-Gordon theory \cite{tien}. The latter represents a simple heuristical treatment of the periodically driven transport, that captures some essential features of this phenomena \cite{hanggi}. Indeed, according to the Floquet theorem the energy level, $E_0$, is split to infinite number of the sub-levels $E_0+n\omega$, generated by the emission (absorbtion) of $n$ photons. Therefore the process can be considered as a multi-photon assisted tunneling, where
$J_n^2(u/\omega)$, represents the probability of absorbing (or emitting) $n$ photons \cite{hanggi}. The question is whether the Tien-Gordon approach can be extended in a simple way to  oscillating tunneling barriers? This is investigated in next Section.

\section{Oscillating tunneling barrier\label{sec6}}

In the same way as the energy level of quantum well, the tunneling barrier can be periodically modulated by an external field, resulting in a periodic modulation of the tunneling coupling, $\Omega_r(t)=\Omega\, w(t)$, Eq.~(\ref{omt}) (we considered here the wide-band limit). One can easily obtain from the semiclassical (Gamow) formula for tunneling rate that
\begin{align}
w(t)=1+\alpha\sin (\omega t)
\label{b10}
\end{align}
where $\alpha \sim{\delta V_0\over V_0}\ll 1$ with $V_0$ are $\delta V_0$ are the barrier hight and the amplitude of  barrier's oscillations.  Then solving Eq.~(\ref{b77}) for $t>0$ with  $\dot E_0(t)=0$ and keeping only the linear in $\alpha$ terms, we find the occupation of quantum well as a function of time. The results are shown in Fig.~\ref{fig33} for $\Lambda =4\Gamma$, $\alpha=0.1$ and two values of frequency, $\omega =0$ (solid line), $\omega =2\Gamma$ (dashed line, red). The panel (a) corresponds to $E_0=3\Gamma$ and the panel (b) corresponds to $E_0=0$ (aligned with the band-center).
\begin{figure}[h]
\includegraphics[width=9cm]{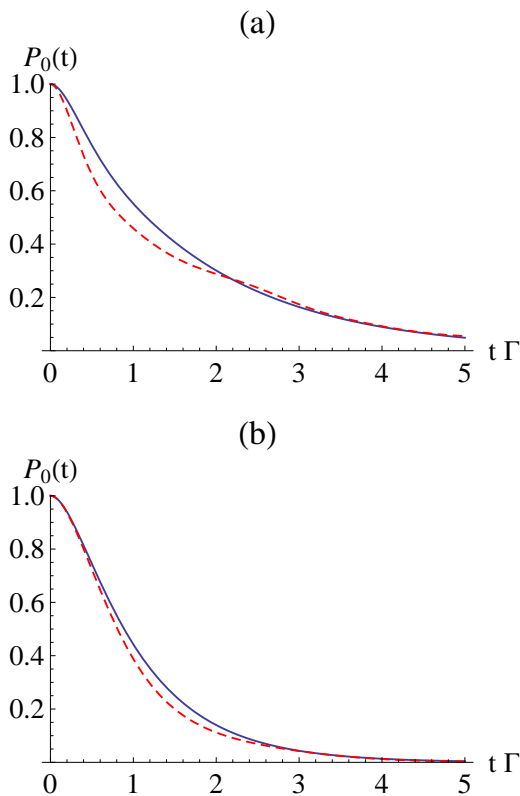}
\caption{(Color online) Probability of survival for oscillating tunneling barrier for $\Lambda =4\Gamma$, $\alpha=0.1$, and $\omega =0$ (solid) or $\omega =2\Gamma$ (dashed, red), where $E_0=3\Gamma$ (a), and $E_0=0$ (b).
\label{fig33}}
\end{figure}

In contrast with the case of oscillating energy level, Fig.~\ref{fig3}, the oscillating barrier, always speeds up the decay to continuum, irrespectively of whether $E_0$ is aligned or misaligned with the band-center. The reason is that the barrier oscillations affect only the tunneling rate, which increases at average.

The most interesting effect of oscillating barrier appears in the energy spectrum of tunneling particle. In the case of oscillating energy level, $E_0$, the latter is split into sub-levels, Eq.~(\ref{c1}), generated by multi-photon emission (absorbtion). Now the level $E_0$ is not-oscillating, but only its width $\Gamma$. The question is whether oscillations of the level-width would generate sub-levels, similar to Eq.~(\ref{c1}).

The problem can be easily resolved in the case of $\Lambda\to\infty$. Then the energy spectrum of the tunneling particle is given by Eq.~(\ref{b9}), where
\begin{align}
{\cal E}_0(t)\,t=E_0t-i {\Gamma\,t\over2}+i{\alpha\,\Gamma\over\omega}(\cos \omega t -1)
 \label{b11}
\end{align}
Using
\begin{align}
\exp\Big[-{\alpha\Gamma\over\omega}\cos(\omega t)\Big]=\sum_{n=-\infty}^{\infty}
I_n\Big({\alpha\Gamma\over\omega}\Big)e^{-i\,n\omega t}
\end{align}
where $I_n(x)$ are the Modified Bessel functions of the first kind, we rewrite Eq.~(\ref{b9}) as
\begin{align}
P_r(t)&={\Gamma\over 2\pi} \left|\int\limits_0^t\big[1+\alpha\sin (\omega t')\big]\right.\nonumber\\
&\left.\times \sum_{n=-\infty}^\infty
e^{i\big(E_r-E_0-n\omega +i{\Gamma\over2}\big)t'}e^{-\xi}
I_n(\xi)dt'\right|^2
\label{b12}
\end{align}
where $\xi=\alpha\Gamma/\omega$. Taking the limit of $t\to \infty$, we finally obtain
\begin{align}
\bar P(E_r)&={\Gamma\over 2\pi}\left|\sum_{n=-\infty}^\infty e^{-\xi}I_n(\xi )\Bigg[{1\over E_r-E_0-n\omega+i{\Gamma\over2}}\right.\nonumber\\
&\left.-{i\,\alpha\, \omega\over \big(E_r-E_0-n\omega+i{\Gamma\over2}\big)^2-\omega^2}
\Bigg]\right|^2
\label{b13}
\end{align}

It is interesting to compare the energy spectrum of tunneling particle, $\bar P(E_r)$, for oscillating energy level, Eq.~(\ref{c1}) with that for oscillating barrier hight, Eq.~(\ref{b13}) for the same values of the corresponding  driving parameters, $u/\Gamma =\alpha$. An example of such  comparison is presented in Fig.~\ref{fig2} for $\alpha =u/\Gamma=.2$, $\omega=2$ and $E_0=0$. The energy spectrum for oscillating energy level is shown by dashed line (red), and for oscillating barrier by solid line. The both curves are rather similar. However, the first Floquete state is more pronounced for oscillating barrier. That is due to the second term of Eq.~(\ref{b13}), generating by a prefactor $w(t)$ in Eqs.~(\ref{endis}), (\ref{b9}).
\begin{figure}[h]
\includegraphics[width=8cm]{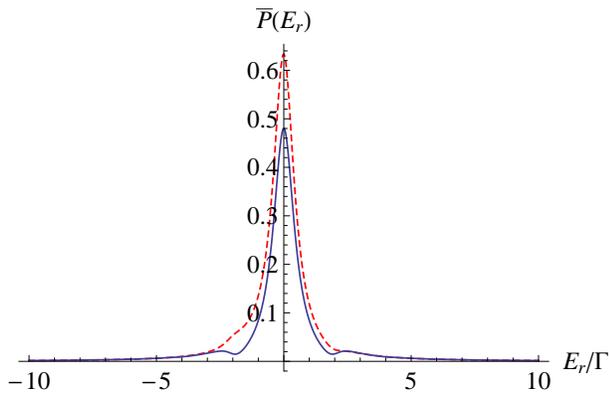}
\caption{(Color online) Energy spectrum of tunneling particle, $\bar P(E_r)$ for oscillating tunneling barrier, Eq.~(\ref{b13}), (solid line), and oscillating energy level, Eq.~(\ref{c1}) (dashed line, red) for $\alpha=u/\Gamma=\omega/\Gamma =0.2$.}
\label{fig2}
\end{figure}

One finds that similar to Eq.~(\ref{c1}), the harmonic oscillations of the  energy-width  manifests the Floquet theorem, as well. As in the previous case, the Floquet sub-levels ($n\omega$) in Eq.~(\ref{b13}) correspond to the multi-photon emission (absorbtion). The difference with the oscillating energy level appears is the probability amplitude for absorbing or emitting of $n$ photons. Now is given by the modified Bessel function, $e^{-x}I_n(x)$ with the argument $x=\alpha\Gamma/\omega$, instead of $J_n(x)$ with the argument $x=u/\omega$. It suggests that a heuristical Tien-Gordon treatment of a periodically driven transport can be extended to oscillating barriers by this replacement.

This can also be understood by using the following arguments. Consider the localized state as a quasi-stationary (Gamow) state of the energy, $E=E_0-i{\Gamma\over2}$. One finds that oscillations of the barrier hight corresponds to oscillations of the imaginary part of the energy, $\Gamma/2$. It implies that the argument of the Bessel function $J(u/\omega)$ in Eq.~(\ref{b13}) must be replaced by $i\,\alpha \Gamma/2$. Using $J_n(ix)= (i)^nI_n(x)$ in Eq.~(\ref{b13}) we recover the first (leading) term of Eq.~(\ref{c1}), up to a factor $\exp (-x\Gamma/\omega)$. The latter cannot be recovered by such simple heuristical argument, but it is necessary for providing correct adiabatic limit for $\omega\to 0$.

\section{Discussion\label{sec7}}

In this paper we studied tunneling from a quantum well to reservoir as continuous Schr\"odinger evolution of the wave-function, localized at $t=0$ inside the well. Solving
the Schr\"odinger equation directly in the time domain, we obtained the wave function in the whole time-interval  $-\infty<t<\infty$, where it is uniquely defined by its value at $t=0$. We emphasize that there is no causality condition ($t>0$) in our approach, in contrast to the Laplace transform method. One needs just to realize that the wave-function in the negative time-interval ($-\infty<t<0$) describes formation of localized state inside the well from extended states of the reservoir. From this point of view we elucidated the old time-arrow problem in the quantum-mechanical context by using our example of tunneling. We hope that our analysis can be useful for future investigations of this important problem.

Transition amplitudes between localized state inside the well and extended states of the reservoir, contain singularity at $t=0$. This singularity has to be taken into account in frequently used Taylor expansion of the evolution operator. In the case of Markovian reservoir, the singularity produces discontinuity in the first derivative of occupation of the well as a function of time. It disappears for a finite band-width of the non-Markovian reservoir, but it still exists in the higher order derivatives.

Finally, we applied our results for tunneling from periodically modulated energy-level of the well and through periodically oscillating tunneling barrier. From our exact expression for the tunneling rate and the energy distribution in the continuum, we easily recover the Floquet theorem for oscillating energy level. The same analysis of our exact expressions allows us to extend the Floquet theorem for oscillating energy width. This allows us to extend the Tien-Gordon approach to the oscillating tunneling barriers, which has not been done before.

Despite the problems, discussed in this paper are quite different, all of them are investigated by using the same treatment, based on the time-dependent approach. We anticipate that this approach can be useful for a variety of different physical problems, related to quantum transport (see for instance \cite{gur,GAE}). Some of them are under current investigation.

\begin{acknowledgements}
The hospitality of the Beijing Normal University and the Beijing Computational Research Center where a part of this work had been done, is gratefully acknowledged. The author is thankful to A. Aharony, O.Entin-Wohlman for very useful discussions.
\end{acknowledgements}


\begin{references}
\bibitem{fn1} The same behavior takes place when the system is not in a pure state, but in a statistical mixture, described by density-matrix.
\end{references}
\end{document}